\documentstyle[epsfig,12pt,subfigure]{article}

\newcommand{\be}{\begin{equation}}
\newcommand{\ee}{\end{equation}}
\newcommand{\bea}{\begin{eqnarray}}
\newcommand{\eea}{\end{eqnarray}}
\newcommand{\bean}{\begin{eqnarray*}}
\newcommand{\eean}{\end{eqnarray*}}
\newcommand{\kkpi}{\mbox{$K^{0}_{S} K^{\pm} \pi^{\mp}$} }
\newcommand{\etapipi}{\mbox{$\eta \pi \pi$} }

\newcommand{\gapproxeq}{\lower
.7ex\hbox{$\;\stackrel{\textstyle >}{\sim}\;$}}
\newcommand{\lapproxeq}{\lower
.7ex\hbox{$\;\stackrel{\textstyle <}{\sim}\;$}}
\begin{document}
\begin{titlepage}
\begin{tabbing}
right hand corner using tabbing so it looks neat and in \= \kill
\> {RAL-97-029}   \\
\> {BHAM-HEP/97-03}   \\
\> {16 June 1997}
\end{tabbing}
\baselineskip=18pt
\vskip 0.7in
\begin{center}
{\bf \LARGE Implications of the Glueball-$q \overline q$ filter on
the $1^{++}$ nonet} \\
\vspace*{0.9in}
{\large Frank E. Close}\footnote{\tt{e-mail: fec@v2.rl.ac.uk}} \\
\vspace{.1in}
{\it Rutherford Appleton Laboratory}\\
{\it Chilton, Didcot, OX11 0QX, England}\\
\vspace{0.1in}
{\large Andrew Kirk}\footnote{\tt{e-mail: ak@hep.ph.bham.ac.uk}} \\
{\it School of Physics and Astronomy}\\
{\it Birmingham University}\\
\end{center}
\begin{abstract}
The application of our glueball-$q\bar{q}$ filter to the centrally produced
$K\bar{K}\pi$ system shows that the $f_1(1285)$ and $f_1(1420)$
have the same behaviour; namely consistent with the
$f_1(1420)$ being the partner to the
$f_1(1285)$ in the $^3P_1$ nonet of axial mesons.
We determine a flavour
singlet-octet mixing angle of $\sim 50^o$
for this nonet and highlight that the existence
of the supposed $f_1(1510)$ needs confirmation.
\end{abstract}
\end{titlepage}
\section{Introduction}
\par
Recently we announced the discovery of a kinematic filter
that separates $q\bar{q}$ and glueball states
in central production\cite{ck97} and illustrated its success in several
channels\cite{WADPT}. This filter has now been applied to the
process $pp \to p(K\bar{K}\pi)p$ where the
prominent states in the $K\bar{K}\pi$ system have
$J^{PC} = 1^{++}$\cite{wakkpi}.
 In the present paper we shall argue
that this new technique establishes the $f_1(1285)$ and $f_1(1420)$
as $q\bar{q}$ states, in contrast to a widespread opinion that the
nonet contains an $f_1(1510)$ while the $f_1(1420)$ is a non-$q\bar{q}$
state\cite{re:FOURQUARK}
or $K^* \overline K$ molecule
\cite{re:WEINSTEIN,LONGACRE}.

We shall show that there is a substantial body of data consistent with
the $D\equiv f_1$ (1285) and $E\equiv f_1$ (1420)
belonging to a $q\bar{q}$ nonet with a consistent mixing

\bea
D \simeq \mid n\bar{n} \rangle - \delta \mid s\bar{s} \rangle \nonumber \\
E \simeq \mid s\bar{s} \rangle + \delta \mid n\bar{n} \rangle \nonumber
\eea
with $\delta \simeq$ 0.4 - 0.5 $(n\bar{n} \equiv \frac{1}{\sqrt{2}} (u\bar{u} +
d\bar{d})$).  The essential feature is the $n\bar{n} (s\bar{s})$ dominance in
$D(E)$ with destructive (constructive) admixture of the other flavour.

We shall also question the existence or interpretation of the supposed
$f_1(1510)$, highlight possible anomalies in the data and
identify critical questions for experimental investigation.

\subsection{A brief history}
Originally
the E/$f_1(1420)$ was thought to be the $s \overline s$ isoscalar member of
the ground state $1^{++}$ nonet, the other members being
the $a_1(1260)$ triplet, the $K_1(1270/1400)$ and the $f_1(1285)$.
The E/$f_1(1420)$ was found to decay dominantly to $K^* \overline K$ hence
reinforcing its $s \overline s $ assignment.
However, problems with this interpretation then began to emerge.
Firstly, it is commonly accepted that
$ s \overline s$ objects should be preferentially produced
in $K^-$ incident experiments but
in the study of the reaction
\begin{center}
$ K^- p$ $ \rightarrow$ \kkpi $\Lambda$
\end{center}
two experiments
\cite{re:GAV,re:LASS}
observed only weak evidence for a E/$f_1$(1420) signal. Instead
they found evidence for a new
J$^{PC}$~=~$1^{++}$ state with a mass of 1.53~GeV and a width of 100 MeV,
called the D$^\prime$/$f_1(1510)$.
It was suggested that this state is a better
candidate for the $ s \overline s$  member of the $1^{++}$ nonet based on the
fact that its production was more compatible with that of
an $s \overline s $ state.
Further evidence for this classification came from studying the
nonet mixing angle from various sources~\cite{re:GAV,OST}
and from a study of hadronic $J/\psi$
decay~\cite{MPP,GE}.
\par
Therefore, the $1^{++}$ nonet appears to have ten members with the
E/$f_1$(1420)
thought to be the extra state.
As a $1^{++}$ state
its mass is generally regarded as too low to be a glueball (lattice
QCD predicts the lightest $1^{++}$ glueball to be at $4.0 \pm 0.5$~GeV
\cite{ukqcd})
and suggestions have been made that it is a four quark state
\cite{re:FOURQUARK}
or $K^* \overline K$ molecule
\cite{re:WEINSTEIN,LONGACRE} (there is however no sign of an $I=1$ partner).
\par
In this paper we will reanalyse the data on the isoscalar $1^{++}$ states.
The new input in this paper are the
data from central production which show that the $f_1(1285)$ and the
$f_1(1420)$ both behave the same with $dp_T$, specifically in the way
that parallels other standard $q \overline q$ states~\cite{WADPT}.
\subsection{Criticism of Evidence for the $f_1(1510)$}
\par
The PDG~\cite{PDG96} cites 4 references for the $f_1(1510)$:
the two original $K^-p$ experiments
\cite{re:GAV,re:LASS},
a $\pi^- p $ experiment~\cite{CHUNG} and
a $\gamma \gamma^*$ experiment~\cite{gg1510}.
The masses and widths observed in each experiment are given in
table~\ref{tab1}.
However, we would highlight some questions about these assignments.

The PDG chooses the values coming from the $\pi^- p$ experiment performed at
the
MPS~\cite{CHUNG}
as the parameters it quotes for the $f_1$(1510).
This choice is rather bizarre since
the peak observed is only at best a 2.5$\sigma$ effect
(see fig.~\ref{fi:chung})
and it does not appear in the $1^{++}$ wave
(see fig.~\ref{fi:chungpwa}b where no significant structure is observed
in the $1^{++}$ wave in the 1.5 GeV region).
The only hint for
a $1^{++}$ assignment comes from a statistically weak phase motion.
We regard it as significant that the same group, in a high statistics
experiment using a $K^-$ beam
subsequently found
no evidence for this state~\cite{CHUNGK}.
\par
The original evidence for the $f_1$(1510) (or as it was originally called,
the D$^{\prime}/f_1(1530)$) had come
from the (low statistics) study of the \kkpi system in
$K^- p$ interactions. Two experiments reported evidence for an axial meson
in the 1.53 mass region though different analysis methods
were needed to extract the signal. The evidence for the state comes from
an observation of an asymmetry in the Dalitz plot of the \kkpi system.
In the one experiment~\cite{re:GAV} this asymmetry is interpreted
as due to incoherent $K^*$ production; in the second
experiment~\cite{re:LASS} it is explained as being due to an interference
between the hypothesised $f_1(1510)$ and the $h_1(1380)$.

The mutual consistency of the two experimental signals is debatable. It is
notable that both experiments suffer
from low statistics; by contrast two subsequent
high statistics studies of the $K^-p$ reaction~\cite{CHUNGK,LEPTONF}
show no evidence for a $1^{++}$ signal in the 1.5 GeV region.
\par
More recent experiments raise further questions concerning the
$f_1(1510)$. No $f_1(1510)$ signal occurs in central production where
the other $1^{++}$ states are clearly seen, nor is it observed in
$p \overline p$ annihilations.
The only other suggestion of a narrow $1^{++}$ state in the 1.5 GeV mass
region comes from BES~\cite{BESKKPI}.
At BES a preliminary partial wave analysis of the reaction
$J/\psi \rightarrow \gamma(\kkpi)$ claims a $1^{++}$ state
in the region of 1.5 GeV. However, we note that this
is in disagrees with results from
from MARKIII~\cite{MARK3} and DM2~\cite{DM2} which had no evidence
for a $1^{++}$ signal in the 1.5 GeV region of  radiative $J/\psi$ decay.
\par
The signal observed in $\gamma \gamma^*$ interactions is in
the $\pi^+\pi^-\pi^0\pi^0$
channel which is bizarre if it is supposed to be an
$s\bar{s}$ state. Furthermore
there is no direct evidence that the state has $J^{P}=1^+$ other than an
unusual
$q^2$ dependence.
We regard as significant the fact that the same group (TPC/2$\gamma$) do not
see any evidence for a state at 1.5 GeV in the \kkpi final
state~\cite{TPC2g}.

There are also theoretical arguments that make it
unlikely that any $1^{++}$ state at $\sim 1510$~MeV could be a
$^3P_1$ $q\bar{q}$ state as at such a mass the width into $K^* \overline K$
would be very
broad~\cite{bcps}. Explicit calculations of the axial meson widths are given in
ref.~\cite{bcps} and are more general than the detailed model.
The empirical feature to note is that the width $a_1 \to \rho \pi$
sets the scale and agrees with the model. The physics is that the
$S$-wave decays rapidly turn on with increasing phase space, making
the widths large and the mesons hard to establish if $M \geq 1450$~MeV.
The width of the candidate $f_1(1285)$ and $f_1(1420)$ are compatible with
the model,
whereas $f_1(1510)$ with ``narrow" width
of $\sim 35\pm15$MeV\cite{PDG96}
would be hard to understand. Thus our point of departure
is to suppose that the $f_1(1420)$ and $f_1(1285)$ are driven by the
$q\bar{q}$ nonet and that the $f_1(1510)$ is either an artifact or a novel
dynamic state.

\subsection{Information from central production}
\par
In central production the $f_1(1285)$ and E/$f_1(1420)$ are clearly
observed~\cite{cenkkpi}. Furthermore they exhibit the same behaviour
as a function of the $dp_T$ filter\cite{ck97,wakkpi}, appearing
sharply when $dp_T > 0.5~$GeV, as do other established $^3P_J$ $q\bar{q}$
states, and vanishing as $dp_T \to 0$. This is consistent with
the $f_1(1420)$ having the same dynamical structure as the $f_1(1285)$, namely
$^3P_1(q\bar{q})$. By contrast, there is no signal for the $f_1(1510)$
at any $dp_T$.

Nor is there any evidence for any $0^{-+}$ contribution in the 1.4~GeV
region. In fact it is interesting to note that $0^{-+}$ states
are suppressed in central production relative to $1^{++}$
states~\cite{cenetapipi}.
Both the $\eta$ and $\eta^{\prime}$ signals are suppressed at small
four-momentum transfers where Double Pomeron Exchange (DPE) is believed
to be dominant~\cite{cenetapipi,cen3pi}. Hence it could be conjectured
that $0^{-+}$ objects do not couple to DPE.
This hypothesis could further be tested by measuring the
cross section of the production of $\eta^{\prime}$ as a function of energy.
The $\eta{\prime}$ cross section has been measured~\cite{WA76UN}
in pp interactions
at an incident beam momentum of 85 and 300~GeV/c and gives
\begin{equation}
\frac{\sigma_{85}(\eta^\prime)}{\sigma_{300}(\eta^\prime)} = 0.2 \pm 0.05
\end{equation}
which is consistent with the $\eta^\prime$ being produced by Reggeon
exchange~\cite{pred}.
\par
This suppression of $0^{-+}$ states in central production has a very important
application since it can help us in untangling other experiments
that see both $0^{-+}$ and $1^{++}$ states. For example,
in the centrally produced \kkpi mass spectrum (see fig.~\ref{fi:cenetapipi}a)
clear signals are observed of the  $f_1(1285)$ and
$f_1(1420)$. However, as can be seen from
fig.~\ref{fi:cenetapipi}b, in the centrally produced
$\eta \pi^+ \pi^-$ spectrum there are clear signals of the
$\eta^{\prime}$ and $f_1(1285)$ but
no signal in the 1.4 GeV region.
Therefore, we can infer that any states
observed
elsewhere prominently in the
1.4 GeV region of the
$\eta \pi \pi$ mass spectrum are {\bf not}
$1^{++}$ and are likely to be pseudoscalar. This fact will
be exploited later when we come to discuss hadronic $J/\psi$ decays.

\section{The $1^{++}$ nonet}

\subsection{Introduction}
In this section we show that the $D\equiv f_1$ (1285) and $E\equiv f_1$ (1420)
form a $q\bar{q}$ nonet with a consistent mixing

\bea
D \simeq \mid n\bar{n} \rangle - \delta \mid s\bar{s} \rangle \nonumber \\
E \simeq \mid s\bar{s} \rangle + \delta \mid n\bar{n} \rangle \nonumber
\label{eq:a}
\eea
with $\delta \simeq$ 0.4 - 0.5 $(n\bar{n} \equiv \frac{1}{\sqrt{2}} (u\bar{u} +
d\bar{d})$).  The essential feature is the $n\bar{n} (s\bar{s})$ dominance in
$D(E)$ with destructive (constructive) admixture of the other flavour.
Expressed in {\bf 1-8} flavour basis we have
\bea
D = \cos\theta \mid 1 \rangle + \sin\theta \mid 8 \rangle \nonumber\\
E = \cos\theta \mid 8 \rangle - \sin\theta \mid 1 \rangle \nonumber
\eea
where $\mid 8 \rangle \equiv \frac{1}{\sqrt{3}} \mid n\bar{n} -\sqrt{2}
s\bar{s}
\rangle$ and we shall find that $\theta \sim 50^o$.

\subsection{Mass formula}
\par\indent
{}From the Gell Mann Okubo Mass formula
\begin{equation}
cos^2(\theta) = \frac{2m_{K_1} + 2m_{K_2} -m_{a_1} - 3m_E}
{3(m_D -m_E)}
\end{equation}
we get $\theta$~=~(52.5$^{+6.7} _{-5.0}$)$^0$ .
It should be noted that this mass formula assumes
that symmetry breaking in the masses is pure
octet; if the $s \overline s$ were initialy at 1480 MeV,
as is predicted in the Godfrey-Isgur
model~\cite{GIM}, then the physical resonance
could be shifted to
the $f_1(1420)$ mass due to the presence of the $K^* \overline K$ threshold
through a mechanism similar to that suggested in ref.~\cite{torn}.
Such a mass shift would imply
$\theta$~=~43$^0$ which highlights the lack of sensitivity of the mass formula
and the artificial reliance on pure $q \overline q$ interpretation.

\subsection{SU(3) coupling formula}
\par\indent
{}From the SU(3) coupling formula~\cite{SU3} an expression can be derived
for the nonet mixing angle in terms of the partial decay rates and
reduced couplings such that
\begin{equation}
cos^2(\theta) = \frac{\Gamma(E \rightarrow K^* \overline K) m^2_E}
{<q_{K^*}> g^2_A}
\end{equation}
where $g^2_A$ is derived from the decay $a_1(1270) \rightarrow \rho\pi$
and gives
$\theta$~=~(63$^{+5} _{-4}$)$^0$.
This calculation assumes SU(3) symmetry for decays involving $s \overline s$
and $n \overline n$ creation. If this constraint is relaxed, $\theta$ can
be shifted considerably in either direction.
\subsection{Radiative decay of $f_1(1285)$}
\par\indent
{}From radiative $f_1(1285)$ decays to a $\phi$ and $\rho$ the nonet mixing
angle
can be calculated from
\begin{equation}
\frac{Br(f_1(1285) \rightarrow \phi \gamma)}{Br(f_1(1285) \rightarrow \rho
\gamma)} = \frac{4} {9} \frac{q^3(f_1 \rightarrow \phi \gamma)}{q^3(f_1
\rightarrow \rho \gamma)}tan^2(\theta - \theta_{ideal})
\end{equation}
where $\theta_{ideal}$~=~35.3$^0$ is the ideal nonet mixing angle.
The PDG~\cite{PDG96} gives
Br($f_1(1285) \rightarrow \phi \gamma$)~=~(8.0~$\pm$~3.1)10$^{-4}$ and
Br($f_1(1285) \rightarrow \rho \gamma$)~=~(6.6~$\pm$~3.1)10$^{-2}$ which
implies
$\theta$~=~(56$^{+4} _{-5}$)$^0$ .
\subsection{$\gamma \gamma^*$ and radiative $J/\psi$ decays}
Here we reassess earlier analyses of $\gamma\gamma^*$ and $J/\psi$ radiative
decays, incorporating our new results from central production.  Recall
that we defined
\bea
D = \cos\theta \mid 1 \rangle + \sin\theta \mid 8 \rangle \nonumber\\
E = \cos\theta \mid 8 \rangle - \sin\theta \mid 1 \rangle \nonumber
\eea
where $\mid 8 \rangle \equiv \frac{1}{\sqrt{3}} \mid n\bar{n} -\sqrt{2}
s\bar{s}
\rangle$.  For simplicity we shall ignore phase space and form factors; these
tend
to cancel out in ratios, do not essentially affect our conclusions and enable
readers to modify according to taste.

The initial input is from radiative $J/\psi$ decays where the
branching ratio of the $J/\psi$ to $\gamma D$ and $\gamma E$ are related by
\begin{equation}
\frac{B(J/\psi\rightarrow\gamma E)}{B (J/\psi\rightarrow\gamma D)} =
\frac{0.83 \pm 0.15}{0.65\pm 0.10} \frac{1}{B(E\rightarrow K\bar{K}\pi)}
\equiv
tan^2\theta
\label{ed1}
\end{equation}
and if we assume that
$B(E\rightarrow K\bar{K}\pi)$~=~1 this gives $\theta$~=~$48.5^{+ 3}_{-3.8}$.
The second input is  from $\gamma \gamma^*$ interactions where
\begin{equation}
\frac{\Gamma(E\rightarrow\gamma\gamma^*)}{\Gamma
(D\rightarrow\gamma\gamma^*)} = \frac{0.34 \pm 0.18}{B(E\rightarrow
K\bar{K}\pi)} \equiv tan^2(\theta - 19.5^0)
\label{ed2}
\end{equation}
and again, assuming
$B(E\rightarrow K\bar{K}\pi)$~=~1 gives $\theta$~=~$49.7^{+ 5.6}_{-8.0}$.
The data used here are from the PDG~\cite{PDG96}
and the $\theta$ dependence is derived in ref.\cite{cfl}.

Independent of $B(E\rightarrow K\bar{K}\pi)$ we can calculate
the ratio of eqs (\ref{ed1}) and (\ref{ed2}) which constrains

$$
 f(\theta) = \frac{tan^2(\theta - 19.5^0)}{\tan^2\theta} = 0.27\pm 0.15
$$
In fig.~\ref{fi:tan} $f(\theta)$ is displayed along with the
mean value and the one sigma bands.
As can be seen,
within one sigma
there are three regions of solution
(a) $12^0 - 14.5^0$ (b) 96.5$\pm 1.5^0$ (c)
$31.5^0 - 78^0$.
However we can eliminate (a) and (b), and tighten (c) as follows

\noindent (a) \quad The fact that $B(E\rightarrow K\bar{K}\pi) \leq 1$ implies,
via eqn(5),
that $B(J/\psi\rightarrow\gamma D) \lapproxeq
B(J/\psi\rightarrow\gamma E) $.
Small values of $\theta$, as in solution (a), imply $\mid D\rangle
\simeq \mid 1 \rangle$,  $\mid E \rangle \simeq \mid 8 \rangle $
and hence
$B(J/\psi\rightarrow\gamma D) \gg B(J/\psi\rightarrow\gamma E) $,
inconsistent with the above.  This is a general result, insensitive to phase
space
or form factors.

\noindent (b) \quad $\theta \simeq 90^0$ implies that
$B(J/\psi\rightarrow\gamma E) \gg
B(J/\psi\rightarrow\gamma D)$.  This would only be attainable if
$B(E\rightarrow  K\bar{K}\pi)   \rightarrow 0$.  However this is unlikely for
the following reasons\\
(i)\quad The data from central production
(section 1.3 and fig.3) show a prominent signal in $E\rightarrow
K\bar{K}\pi$ and no presence in  $E\rightarrow \eta\pi\pi$ which suggests
$B(E\rightarrow  K\bar{K}\pi)$ is not small.\\
(ii) \quad The quark model calculations of ref.\cite{bcps} expect that for
$1^{++}$
at this mass both $\mid s\bar{s}\rangle$ and  $\mid n\bar{n}\rangle$
dominantly couple to $KK^*$.  The phases with $\theta \simeq 90^0$ give
constructive contribution for $E\rightarrow K\bar{K}\pi$ which argues for
$B(E\rightarrow  K\bar{K}\pi)\rightarrow 1$ .  These general conclusions are
insensitive to phase space and form factors.

\noindent (c)\quad The solutions cover the range
\bea
D &=& 0.84\mid 1 \rangle + 0.54 \mid 8 \rangle \qquad \theta =
31.5^0\nonumber\\
D &=& 0.34\mid 1 \rangle + 0.94 \mid 8 \rangle \qquad \theta =
78^0\nonumber
\eea
However, the lower end of this range is eliminated by
$$
\frac{B(J/\psi\rightarrow\gamma E)}{B (J/\psi\rightarrow\gamma D)}
\gapproxeq
\frac{0.9}{B(E\rightarrow K\bar{K}\pi)}
$$
at $1\sigma$,
which restricts the mixing angle $\theta \gapproxeq 40^0$ (the solution in
ref.~\cite{cfl} corresponds to $\theta = 55^0$).
\par
Although the mixing in the singlet-octet basis appears to be
constrained somewhat imprecisely,
the qualitative and robust feature of
the solutions is more transparent in the flavour basis
\bea
D = 0.98\mid n\bar{n} \rangle - 0.14 \mid s\bar{s} \rangle \qquad \theta =
40^0\nonumber\\
D = 0.82\mid n\bar{n} \rangle - 0.56 \mid s\bar{s} \rangle \qquad \theta =
78^0\nonumber
\eea
which shows the dominance of  $\mid n\bar{n} \rangle$ and negative phase
relative to $\mid s\bar{s} \rangle$.  Qualitatively it is this destructive
phase that
reduces $J/\psi\rightarrow\gamma D$ relative to $J/\psi\rightarrow\gamma E$
due to $\langle gg\mid n\bar{n}\rangle$ fighting  $\langle gg\mid
s\bar{s}\rangle$ in $J/\psi\rightarrow\gamma gg  \rightarrow\gamma (D,E)$.
The absolute values will be affected slightly by phase space and form factors
but
these generic features are robust.  The $\theta \rightarrow 40^0$ correlates
with
$B (E\rightarrow K\bar{K}\pi) \rightarrow 1$, whereas
the   $\theta \rightarrow 78^0$
has $B (E\rightarrow K\bar{K}\pi) <  0.1$.  The $n\bar{n}$ dominance
will cause
$$
B (J/\psi\rightarrow D \omega(n\bar{n})) > B (J/\psi\rightarrow D\phi
(s\bar{s}))
$$
(we discuss this later).

The $\gamma\gamma$ widths also impose a constraint where from eqn (\ref{ed2})
the larger $\theta$ correlate with small $B (E\rightarrow K\bar{K}\pi)$.  The
data from central production suggest that this
branching ratio is large; the quark model analysis\cite{bcps}
of quasi-two body decays suggests that this will indeed be near 100\%.
\subsection{Solving the hadronic $J/\psi$ problem}
\par
It is interesting that these various data are independently consistent with
$\theta \sim 50^o$ whereby

\bea
D \simeq \mid n\bar{n} \rangle - \delta \mid s\bar{s} \rangle \nonumber \\
E \simeq \mid s\bar{s} \rangle + \delta \mid n\bar{n} \rangle \nonumber
\eea
with $\delta \simeq$ 0.4 - 0.5. We now examine what has been a
cause for the uncertainty in the assignment of the
$f_1(1420)$ as the $s \overline s$ member of the $1^{++}$ nonet, namely
the claim that it is seen opposite the
$\omega$ and not the $\phi$ in hadronic $J/\psi$ decays.
In this section we shall discuss the basis on which this claim is made.
\par
In hadronic $J/\psi$ decays a signal is observed opposite the $\omega$
in the 1.4~GeV region of both the \kkpi and \etapipi mass spectrum.
A spin analysis favours $J^{PC}$~=~$1^{++}$ for both states.
In the \kkpi channel the structure has a mass of 1438~$\pm$~4~MeV
 which is 3 $\sigma$ higher than the
mass of the $f_1(1420)$, and
a width of 94~$\pm$~12 MeV.
The state is also found to decay to $K^* \overline K$
and $a_0(980)\pi$ in the ratio 2:1 which is in contradiction
to the dominance of the $K^* \overline K$ decay mode found for the
$f_1(1420)$ in radiative $J/\psi$ decays and in central production.

Firstly it should be remembered that the hadronic analysis on which this claim
is based was performed prior to the discovery that the $\iota$
peak observed in radiative $J/\psi$ decays has a detailed infrastructure.
Secondly from
the lack of a $f_1(1420)$ signal in the \etapipi mass spectrum of
central production (fig. 3) we would claim that the state observed in the
\etapipi mass spectrum opposite the $\omega$ is not the $f_1(1420)$ and
we conjecture that
any signal seen in the 1.4 GeV region of the $\eta \pi \pi $ mass
spectrum must be pseudoscalar.
This then casts doubt on the $1^{++}$ assignment of the peak in the
1.4~GeV region of the \kkpi mass spectrum.
\par
We now consider further checks that can be made of this nonet structure.  As
$\theta\rightarrow 78^0$ and ignoring phase space we would expect as a
minimum that
$$
\frac{B(J/\psi\rightarrow \omega D)}{B(J/\psi\rightarrow \phi D)} \gapproxeq 2
$$
assuming that $\omega \equiv n\bar{n}, \phi \equiv s\bar{s}$.  If $\theta
< 78^0$, or including phase space, this ratio rises; conversely if $\omega$
or $\phi$ are not ideal flavour states, the ratio will fall.
The current best estimate for this value is 2.7~$\pm$~1.8 from
ref.~\cite{wermeskopke}.
This ratio should be measured
more precisely at Beijing or a future $\tau$-charm factory
with special care to
ensure separation of a possible $\eta$(1295) signal from the
$f_1$(1285) ($D$) in
the $\eta\pi\pi$ channel.  An analysis of the
$K\bar{K}\pi$ channel involves more detailed combinatorics
in the $\phi K\bar{K} \pi$ but may enable $\omega D/\omega E$ to be
extracted.  Phase space and form factor effects tend to cancel in the $\phi
D/\omega E$ ratio.  These questions have taken on renewed interest in  view of
recent work on the presence of higher
(multiquark) components in light mesons \cite{bk}.

\subsection{Criticism of claims
for the non-$q \overline q$ nature of the $f_1(1420)$}
\par
The most striking claim for the non-$q \overline q$ nature of the $f_1(1420)$
has been that
in spite of its dominant decay to $K^* \overline K$ it is not
produced strongly in
$K^- p$ interactions where $s \overline s$ states are usually seen.
However, this claim is refuted by the LEPTON-F experiment which has
studied $\pi^-p \rightarrow K^+ K^- \pi^0 n$ and
$K^-p \rightarrow K^+ K^- \pi^0 Y$~\cite{LEPTONF}.
In the $\pi^-p$ reaction they have a strong $f_1(1285)$ signal but there
is no statistically significant structure in the 1.4~GeV region,
from which they infer that
\begin{equation}
tan^2(\theta - \theta_{ideal}) =
\frac{\sigma(\pi^-p \rightarrow f_1(1420)n)}{\sigma(\pi^-p \rightarrow
f_1(1285)n)} < 0.05
\end{equation}
and hence $\theta < 48^o$. This is
consistent with a dominant $s \overline s$ content in
the $f_1(1420)$ and with our foregoing analyses. Furthermore, in the reaction
$K^-p \rightarrow K^+ K^- \pi^o Y$ they observe a clear signal in the
$f_1(1420)$ region from which they calculate that
\begin{equation}
\frac{K^-p \rightarrow f_1(1420)Y}{\pi^-p \rightarrow f_1(1420)n} > 10.
\end{equation}
again consistent with a dominant $s \overline s$ content.
\par
In addition, one of the early $K^-p$ experiments~\cite{re:GAV}
which claimed a signal for the $D^{\prime}/f_1(1530)$ also claimed
a peak in the 1.42~GeV region consistent with being the
$f_1(1420)$. Therefore, in summary, it is not so simple to claim that
the $f_1(1420)$ is not seen in $K^-$ incident experiments.
\par
Fig.~\ref{fi:sum} shows a summary of the derivation of the singlet-octet
mixing angle from the different methods discussed above,
assuming that the $f_1(1285)$ and the $f_1(1420)$ are the
isoscalar members of the nonet.
As can be seen the angle calculated from the different methods
are consistent and give an average value of 53$^0$
(excluding the LEPTON-F upper limit).
It should be noted that the errors shown in fig.~\ref{fi:sum}
represent the experimental error on the measured quantities only.
Errors due to the theoretical uncertainty in the
derivation of the formulae used can only help to make the
values more consistent.
\section{Outstanding problems and Summary}
\par
The major problem is to determine the number of $1^{++}$ states.
In summary there is no doubt that the $f_1(1285)$ and $f_1(1420)$ exist.
Without the existence of the $f_1(1510)$ the only major problem
with describing the $f_1(1420)$ as the (dominantly)
$s \overline s $ nonet member
has been the possibility
that it is not produced copiously in $K^- p $ reactions, however,
this claim would be denied by at least one experiment~\cite{LEPTONF}.
A summary of the $f_1(1510)$ is that two $K^- p$ experiments have weak
evidence for it, and two others do not see it. One $J/\psi$ experiment
may see it while two do not and there is a small possibility that
it is observed in $\pi^- p$ reactions.
\par
The other major uncertainty is in the $J/\psi$ hadronic decays.
A reanalysis of the $\iota$ region is needed allowing for modern
insights into its detailed infrastructure and our conjecture
on the role of $0^{-+}$ and $1^{++}$ in this region.

In $J/\psi$ radiative decay each experiment has a different
interpretation~\cite{BESKKPI,MARK3,DM2}.
The analysis
of ref.\cite{cfl}
also may have relevant constraints in this regard.
These questions should be reexamined at BES and at a
future $\tau$-charm factory.

Data on $K^-p \to K_s^0K_s^0 \pi^0 Y$ can access $C=+$ and hence eliminate
uncertainties involving possible $C=-$ content
that have clouded some analyses~\cite{re:LASS,CHUNGK}.
However since such data are unlikely in the immediate future,
what should be performed is
a reanalysis of the MPS data since there is an inconsistency in them;
in $\pi^-$ incident a signal for the $f_1(1510)$ is claimed while
in $K^-$ incident there is no signal.

The conclusion reached from the analysis presented in this paper
is that without confirmation of the existence of the
$f_1(1510)$ the isoscalar members of the $J^{PC}$~=~$1^{++}$ nonet
should be considered to be the $f_1(1285)$ and $f_1(1420)$ with a singlet-octet
mixing angle of approximately 50$^0$. Furthermore, any prominent states in
the $1.4$ GeV region of $\eta \pi \pi$ are unlikely to be
$1^{++}$ and are likely
to be $0^{-+}$.

\newpage
{ \large \bf Tables \rm}
\begin{table}[h]
\caption{Parameters of resonances in the fit to the
$\pi^{+}\pi^{-}\pi^{+}\pi^{-}$ mass spectrum.}
\label{tab1}
\vspace{1in}
\begin{center}
\begin{tabular}{|c|c|c|c|c|c|} \hline
 & & & & & \\
 Reaction&Ref &Mass (MeV) & Width (MeV) &Observed & $J^{PC}$\\
 & & & &decay mode & determined\\
 & & & & & \\ \hline
 & & & & & \\
 $K^- p$ & \cite{re:GAV} &1526 $\pm$ 6 & 107 $\pm$ 15 & $K^* \overline K$ &Yes
\\
 & & & & & \\ \hline
 & & & & & \\
 $K^- p$ & \cite{re:LASS} &1530 $\pm$ 10 & 100 $\pm$ 40 &$K^* \overline K$ &Yes
 \\
 & & & & & \\ \hline
 & & & & & \\
 $\pi^- p$ & \cite{CHUNG} &1512 $\pm$ 4 & 35 $\pm$ 15 &$K^* \overline K$ &No
\\
 & & & & & \\ \hline
 & & & & & \\
 $\gamma \gamma^*$ & \cite{gg1510} &1525 $\pm$ 25 & 200 $\pm$ 50
&$\pi^+\pi^-\pi^0\pi^0$ &No  \\
 & & & & & \\ \hline
\end{tabular}
\end{center}
\end{table}
\newpage
{ \large \bf Figures \rm}
\begin{figure}[h]
\caption{The \kkpi mass spectrum
from ref.~\cite{CHUNG}.
The full circles are for 0.0~$\leq$~-t~$<$~1.0~GeV$^2$/c$^2$
and the open circles are for
0.45~$\leq$~-t~$<$~1.0~GeV$^2$/c$^2.$
}
\label{fi:chung}
\end{figure}
\begin{figure}[h]
\caption{The results of the partial wave analysis
from ref.~\cite{CHUNG}.
a) The $0^{-+}$ wave,
b) the $1^{++}$ wave,
c) the $1^{+-}$ wave and
d) the background (phase space) contribution.
}
\label{fi:chungpwa}
\end{figure}
\begin{figure}[h]
\caption{a) The \kkpi and b) the $\eta \pi^+ \pi^-$ mass spectrum from central
production~\cite{cenetapipi}.
}
\label{fi:cenetapipi}
\end{figure}
\begin{figure}[h]
\caption{The function $\frac{tan^2(\theta-19.5^0)}{tan^2\theta}$
with the 0.27~$\pm$0.15 region shown.
}
\label{fi:tan}
\end{figure}
\begin{figure}[h]
\caption{Summary of the $1^{++}$ singlet-octet mixing angle.
}
\label{fi:sum}
\end{figure}
\newpage
\begin{figure}[ht]
\begin{center}
\epsfig{figure=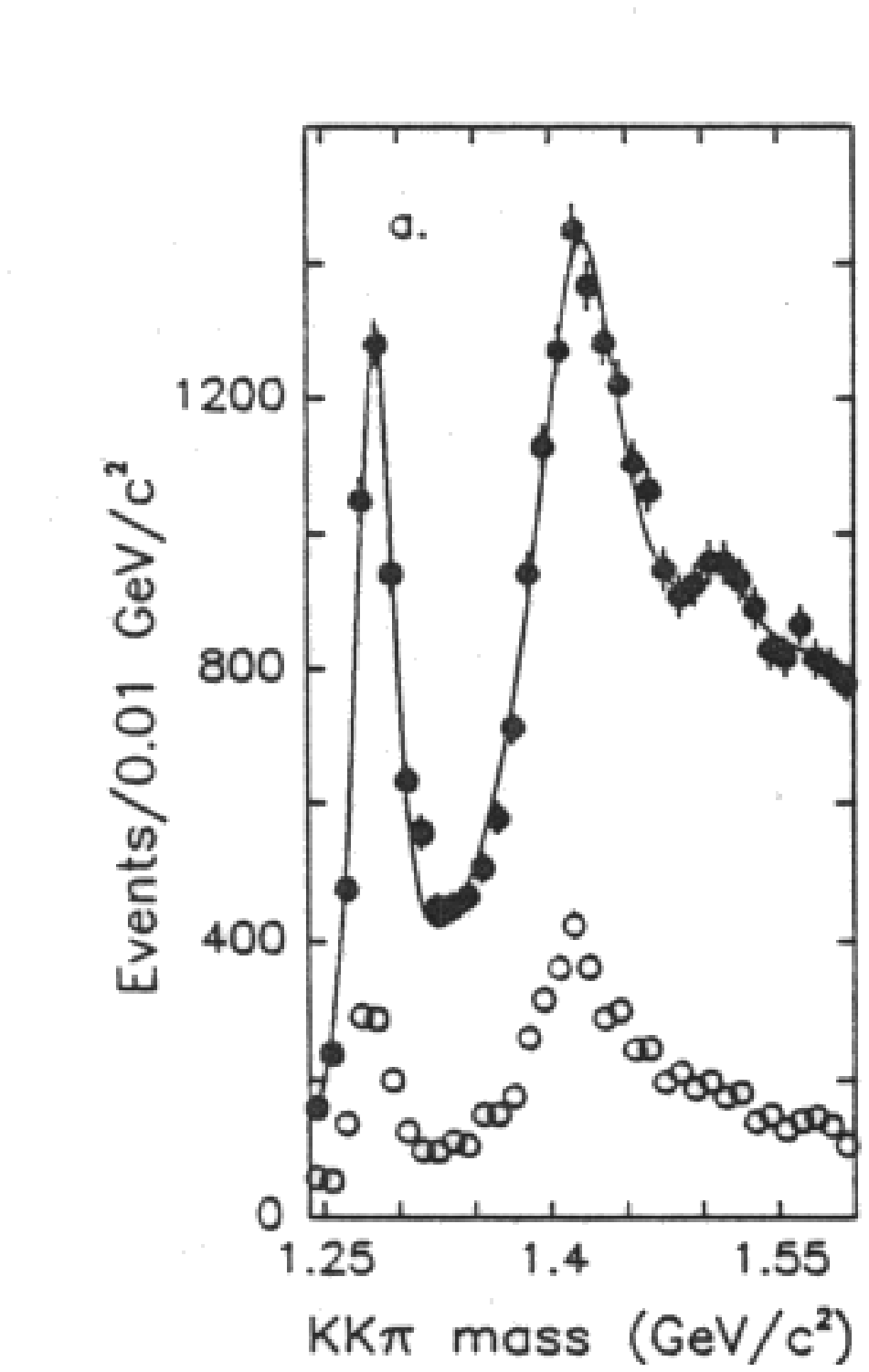}
\end{center}
\begin{center} {Figure 1} \end{center}
\end{figure}
\newpage
\begin{figure}[ht]
\begin{center}
\epsfig{figure=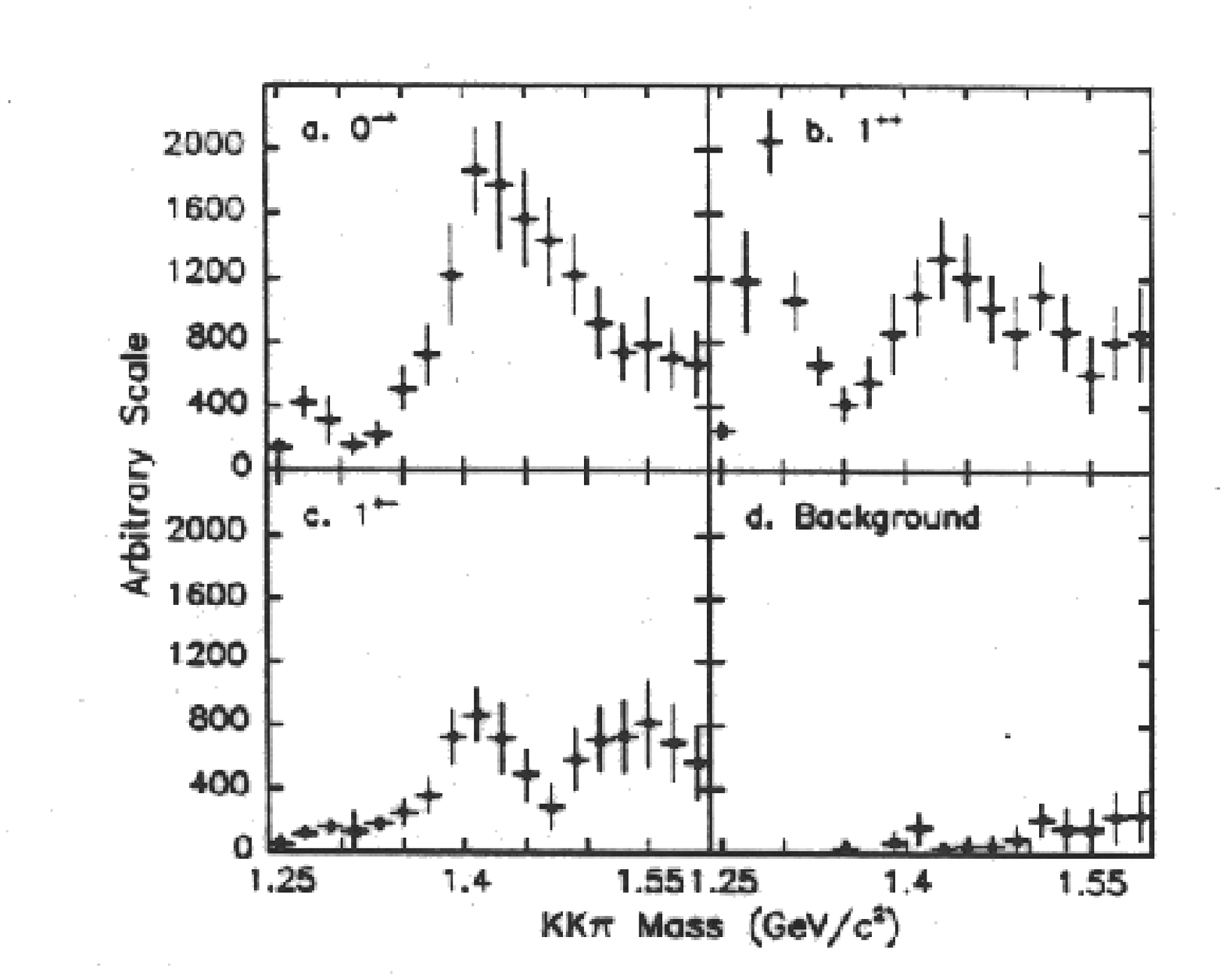}
\end{center}
\begin{center} {Figure 2} \end{center}
\end{figure}
\newpage
\begin{figure}[ht]
\begin{center}
\epsfig{figure=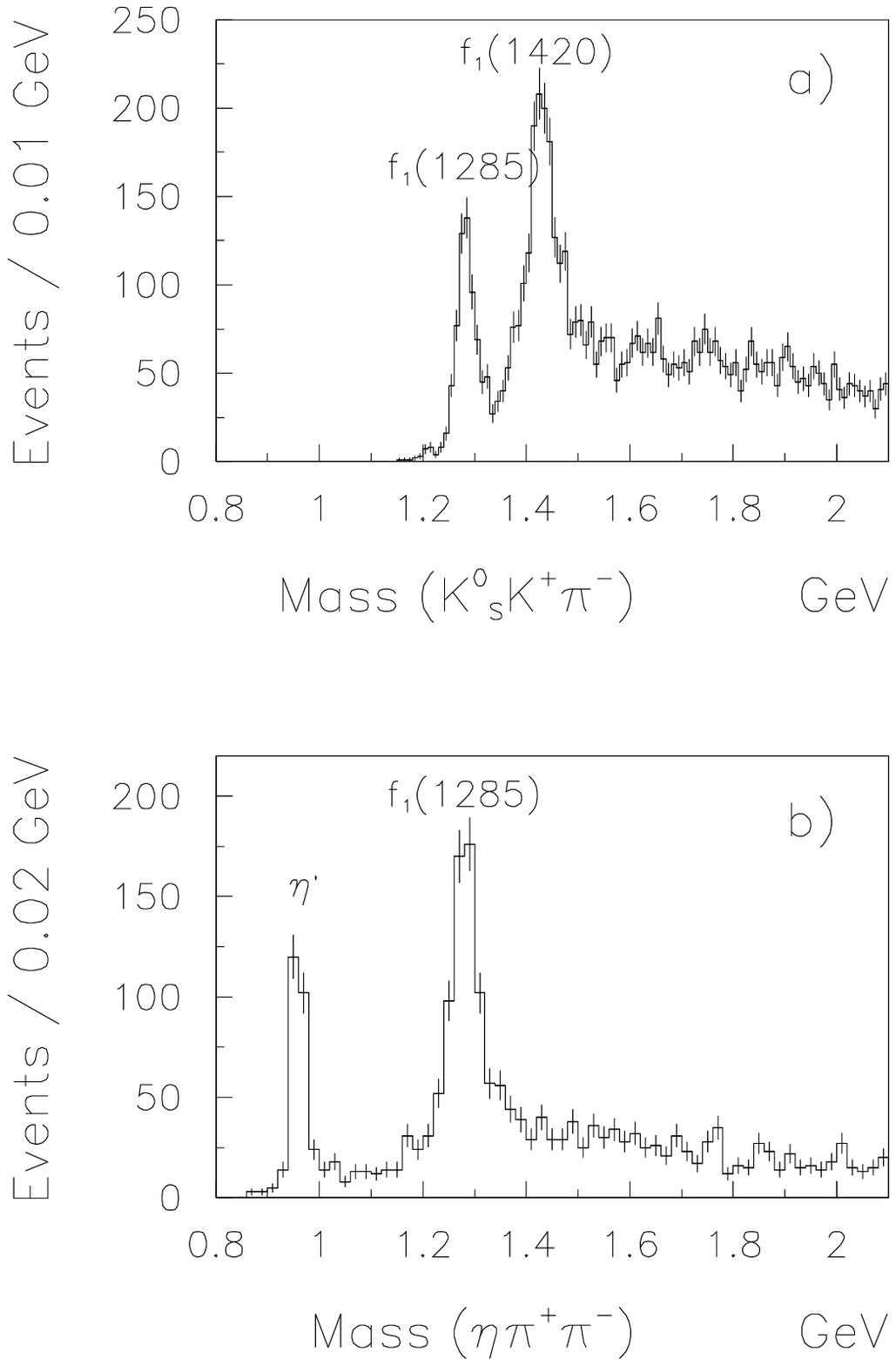}
\end{center}
\begin{center} {Figure 3} \end{center}
\end{figure}
\newpage
\begin{figure}[ht]
\begin{center}
\epsfig{figure=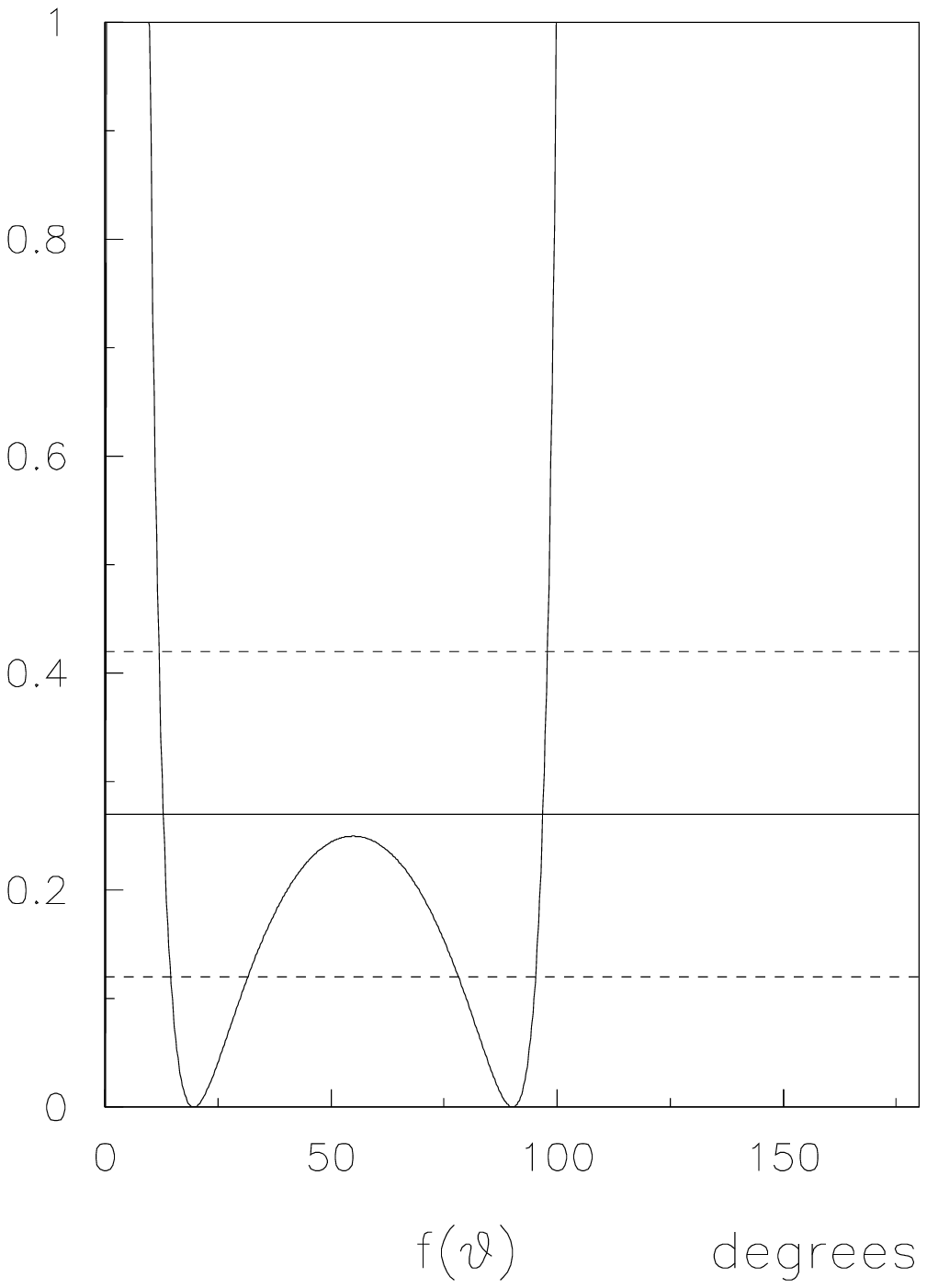}
\end{center}
\begin{center} {Figure 4} \end{center}
\end{figure}
\newpage
\begin{figure}[ht]
\begin{center}
\epsfig{figure=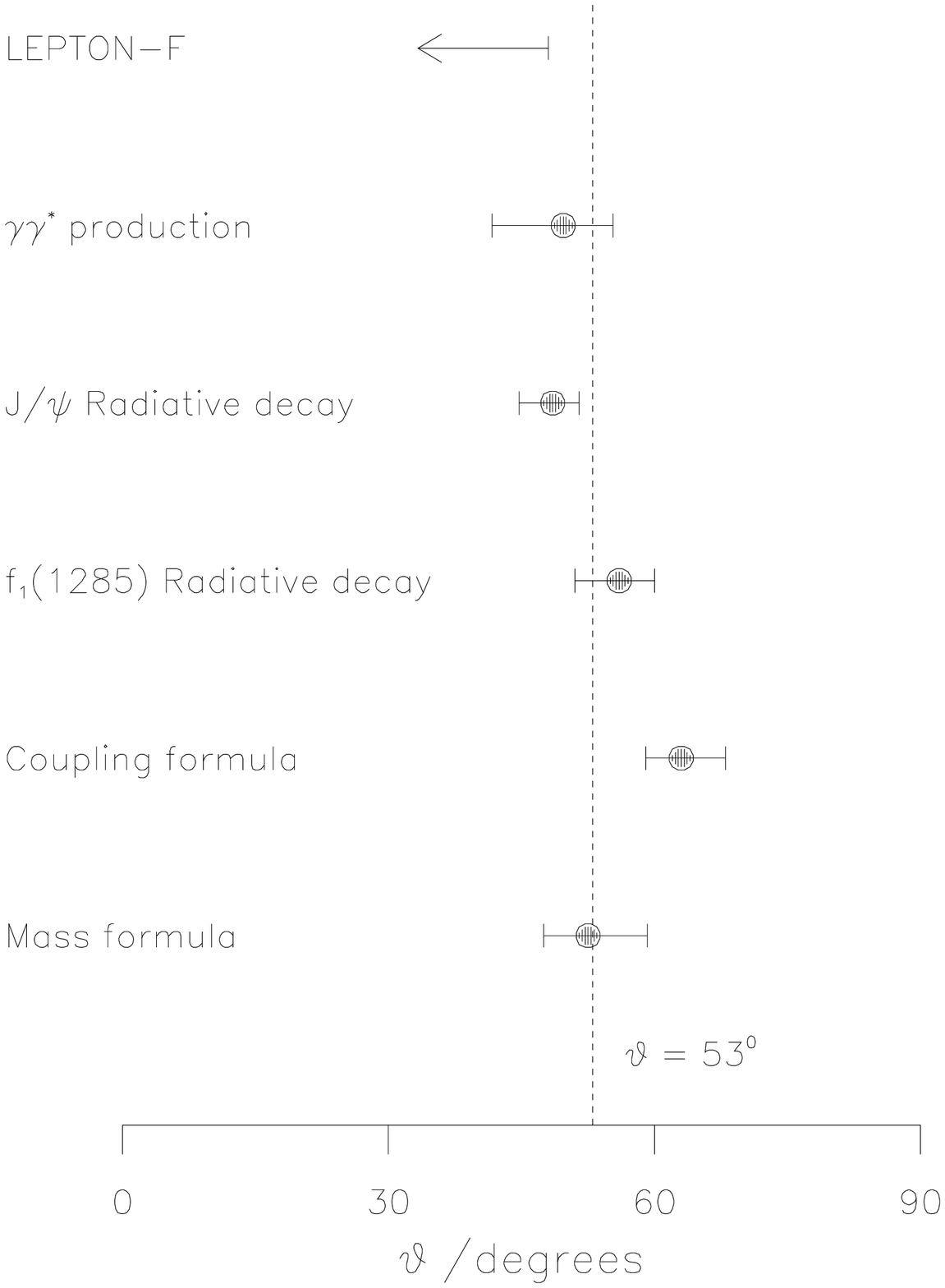,height=17cm,width=17cm}
\end{center}
\begin{center} {Figure 5} \end{center}
\end{figure}
\end{document}